# A 97mW 110MS/s 12b Pipeline ADC Implemented in 0.18μm Digital CMOS

Terje N. Andersen, Atle Briskemyr, Frode Telstø, Johnny Bjørnsen, Thomas E. Bonnerud, Bjørnar Hernes, Øystein Moldsvor

Nordic Semiconductor, Trondheim, Norway

**Abstract**

*A 12 bit Pipeline ADC fabricated in a 0.18μm pure digital CMOS technology is presented. Its nominal conversion rate is 110MS/s and the nominal supply voltage is 1.8V. The effective number of bits is 10.4 when a 10MHz input signal with $2V_{P-P}$ signal swing is applied. The occupied silicon area is $0.86mm^2$ and the power consumption equals 97mW. A switched capacitor bias current circuit scale the bias current automatically with the conversion rate, which gives scaleable power consumption and full performance of the ADC from 20 to 140MS/s.*

## 1. Introduction

CMOS technologies moves steadily towards finer geometries, which provide higher digital capacity, lower dynamic power consumption and smaller area. This results in integration of whole systems, or large parts of systems, on the same chip (System on Chip, SoC). For such systems the key parameters for the building blocks are often low power dissipation and small silicon area. Analog to Digital Converters (ADC) are important building blocks in many of the SoC applications. Therefore, ADCs provided as Intellectual Property (IP) blocks goes into a wide range of applications, spanning from imaging to ultrasound and communication systems. This means that the ADC has to occupy small area and have low power dissipation. Further, it has to be designed in the same technology that the rest of the system. This means that the power supply voltage is low and subsequently the maximum signal swing is lowered. Thus, it is necessary to keep the noise and distortion of the ADC itself low. Additionally, to minimize the cost of the total SoC system, the IP block should be implemented in a pure digital process, avoiding use of costly additional analog process options.

This paper addresses these challenges by presenting a 12 bit pipeline ADC which utilizes a 1.8V supply voltage and is fabricated in a 0.18μm pure digital CMOS technology. In section 2 the used pipeline architecture is presented and section 3 explains the design of some key building blocks of the ADC. Section 4 presents the major measurement results, while section 5 concludes the paper.

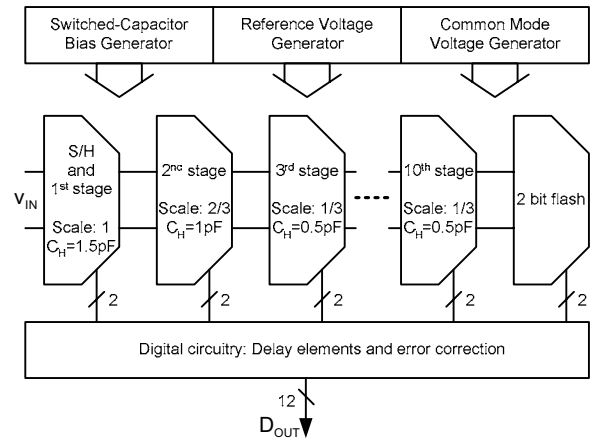

Fig. 1. The pipeline architecture.

## 2. Pipeline Architecture

The pipeline chain is shown in Fig. 1. The chain consists of 10 succeeding 1.5bit stages and a 2bit flash at the end. The input signal is applied directly to the 1st stage, which also performs sample-and-hold. Due to the X2-gain in each stage, the 1st stage has the highest specifications with respect to noise, incomplete settling and distortion. Since the input switches do not utilize bootstrapping, careful design of these is necessary to obtain low distortion. Further, large sampling capacitors ($C_H$) and high bias currents are necessary to keep the noise low and the incomplete settling within the specifications. Due to relaxed requirements of the stages along the pipeline chain these stages are scaled down regards to $C_H$ and bias currents [1]-[2], the 2nd stage with a factor 2/3 and the rest of the stages with 1/3. This result in lower area and lower power consumption with only small degradation in converter performance. The digital output of each stage is passed to a digital circuit, which perform delay and error correction before the digital value appears at the output $D_{OUT}$. The error correction utilizes the half bit of redundancy in each pipeline stage and corrects for errors in the Analog to Digital Sub-Converter (ADSC) contained in the stage. The operation of ADSC is further explained in section 3 together with the rest of the stage architecture.

The reference voltages, Common-Mode (CM) voltages and the bias currents are supplied to the pipeline chain from on-chip circuitry. The reference voltages are derived from the band-gap voltage and are decoupled by off-chip capacitors. The bias current generator is a Switched-Capacitor (SC) circuit which automatically scale the bias currents to the stages with the conversion rate and the absolute value of an on-chip capacitor. The SC bias current generator is also explained in the next section.

## 3. Design of Key Building Blocks

The construction of the pipeline stages is equal for all stages and is shown in Fig. 2 [1]. $V_{INP}^i$-$V_{INN}^i$ is the input signal for $i=1$ and the residue from the preceding stage for $i>1$. The operation of the stage can be divided in two phases. In the tracking phase $\varphi_1$, $V_{INP}^i$-$V_{INN}^i$ is applied over the sampling capacitor $C_H$, which is the parallel connection of the parasitic metal capacitors $C_1$ and $C_2$. When $\varphi_{1B}$ goes down, the sampling switch $S_{1B}$ opens and a sample of $V_{INP}^i$-$V_{INN}^i$ is stored at $C_1$ and $C_2$. At the same time $V_{INP}^i$-$V_{INN}^i$ is also sampled by the ADSC. Further, $\varphi_1$ goes down and $\varphi_2$ goes high and the switches $S_1$ are opened and $S_2$ are closed. The opamp, a two-stage Miller opamp with a differential-pair output stage [3], is now connected in closed loop and the stage is in its amplification phase. ADSC resolves the input sample and pass its digital value to the Decoder and Switching Block (DSB). Dependent on the digital value, DSB connects the reference voltages ($V_{REFP}$ and $V_{REFN}$) or the CM-voltage to the top of $C_1$ and the output $V_{OUTP}^i$-$V_{OUTN}^i$ settles towards the residue. The residue is sampled by the following stage at the end of the amplification phase. At this time instance, the digital output $D_{OUT}^i$ is valid and clocked in to the digital circuit.

Because of the low supply voltage it was necessary to pay the switches special attention. The transmission gates $S_1$ and $S_2$ experience large voltage swings. This results in large transistors to keep the on-resistance of the transmission gate low enough to fulfil the settling requirements. It is especially the PMOS transistor that becomes large due to its lower mobility compared to the NMOS. Therefore, both $S_1$ and $S_2$ use bulk switching of the PMOS transistor [1]. This means that when the switch is on, the bulk (N-well) of the PMOS transistor is switched to source. This results in lower threshold voltage and subsequently lower on-resistance of the transistor. When the switch is off, the bulk is connected to the positive supply voltage, increasing the off-resistance due to increased threshold voltage. The sampling switch $S_{1B}$ is connected to $V_{CM}$ and does not experience large voltage swings. Thus, $S_{1B}$ consist of NMOS transistors only, providing low on-resistance and small parasitic capacitances at the inputs of the opamp.

To obtain low power consumption and low silicon area several actions were taken. First, scaling of the pipeline stages reduce both area and power consumption. Further, in pipeline converters it is common to use non-overlap clocking to ensure that the switches $S_2$ is not closed before $S_1$ is opened. In this design the non-overlap clocking is removed and the sequential operation of the switches is ensured by generating these clocks locally in each stage, as Fig. 3 shows. Removing the non-overlap means that the stage has longer time to settle and the gain-bandwidth of the opamp can be lowered, which further results in lower power consumption.

In modern CMOS technologies the spread in the absolute value of capacitors is large. Instead of large fixed bias currents in the opamp that can handle the largest possible capacitive load, the bias currents in this design are made dependent on the absolute value of the capacitances. Fig. 3 shows the SC bias current generator. The OTA, which is connected in unity gain, ensures that the voltage on its output node BIAS is approximately equal to $V_{BIAS}$. $V_{BIAS}$ is taken from the band-gap voltage circuit and is near independent of variations in process parameters, temperature and supply voltage. The current through the OTA output transistor M0 is set up by $V_{BIAS}$ and the load seen from the node BIAS to ground. This load is the equivalent resistance seen into the SC-circuit. The SC-circuit is clocked by the system clock (CLK) at the conversion rate frequency $f_{CR}$. Further, the current through M0 is mirrored to $I_{BIAS}^1$ to $I_{BIAS}^{10}$, which are applied to stage 1 to 10, respectively. The value of these currents are given by (1) (next page), which shows that the bias current depends on the absolute value of the capacitor $C_B$. Thus, by this technique the nominal bias currents in the opamp can be set lower than in a conventional design and the power consumption is reduced.

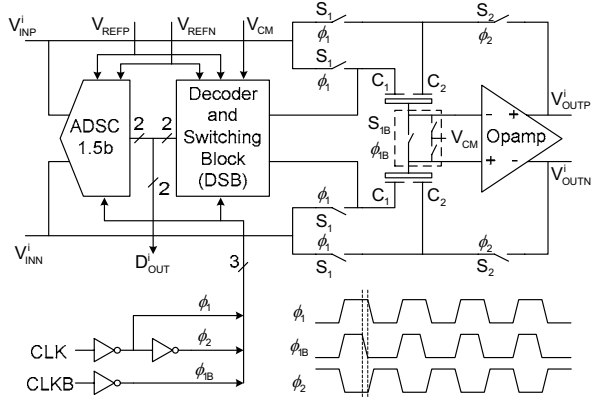

Fig. 2. The pipeline stage.

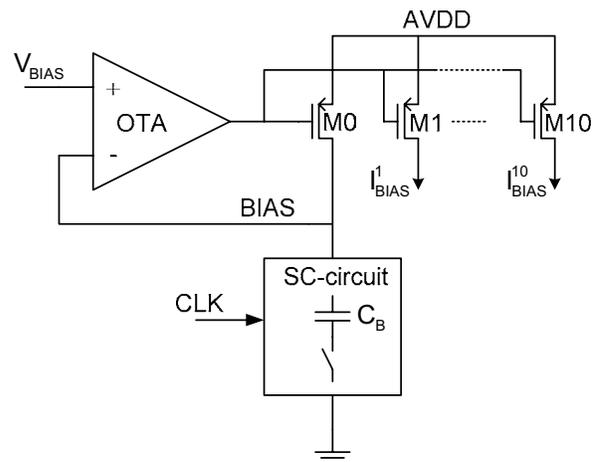

Fig. 3. The SC-bias current circuit.

Further, (1) shows that the bias currents to the pipeline stages are also linearly dependent on $f_{CR}$. This means that the bias currents, and subsequently the power consumption of the ADC, are automatically scaled with the applied clock frequency. This is an important feature in an ADC made as an IP-block due to the wide range of possible applications that this block should fit in to. The effect of the power scaling versus conversion rate is shown in the next section.

$$I_{BIAS}^{i} = C_B \cdot f_{CR} \cdot V_{BIAS} \qquad (1)$$

## 4. Measurement Results

Fig. 4 shows the power dissipation of the ADC (exluding output drivers) versus conversion rate. As predicted by (1) the bias currents, and subsequently the power dissipation, is linearly scaled versus conversion rate. The plot shows a power dissipation of 97mW at 110MS/s and 110mW at 130MS/s.

The dynamic measurements were done by using RF-sources for the input signal and the clocking of the ADC. Both where filtered using high order passive band-pass filters around the applied frequency to remove harmonics and white noise produced by the sources. Fig. 5 shows measurements of SNR, Signal-to-Noise-and-Distortion Ratio (SNDR) and Spurious-Free-Dynamic-Range (SFDR) versus conversion rate. At 110MS/s, SNR and SNDR equal 67.1dB and 64.2dB, respectively. Further, the plot shows that SNDR is above 64dB from 20MS/s up to 120MS/s and is above 62dB (equals 10 effective number of bits) up to 140MS/s. SFDR is above 69 dB from 5MS/s up to 140MS/s. The signal frequency was 10MHz for these measurements.

Fig. 6 shows measurements of SNR, SNDR and SFDR versus input frequency at a conversion rate of 110MS/s. SNR remains above 66dB up to 100MHz. Above 100MHz, jitter is the main noise contribution and SNR is falling with increasing input frequency. SNDR is larger than 60dB up to 40MHz and is thereafter falling due to decreasing SFDR. The reason why SFDR, and subsequently SNDR, are falling off at high input frequencies is the nonlinearity introduced by the input switches of the ADC. These switches are transmission gates where both the channel resistance and the parasitic capacitances are nonlinear. This problem can be solved by using bootstrapping, but this is not done in this design due to potential lifetime issues. The measurements presented in Fig. 5 and Fig. 6 are done with signal amplitude near full scale ($2V_{P-P}$).

In Fig. 7 (next page) the die photograph of the circuit is shown. The layout became compact due to many factors such as strapping the power routing in all metal layers, scaling the power routing for low-power areas and routing above active area. The key measurement results are summarized in Table I (next page). Additional information here is that DNL and INL are ±1.2 LSB and -1.5/+1 LSB, respectively, and the ADC silicon area is $0.86mm^2$.

Equation (2) (next page) is the Figure of Merit (*FM*) given in [4] adjusted such that also the ADC area is included. In (2) *ENOB* is the effective number of bits of the ADC including distortion, $f_{CR}$ is the conversion rate, *A* the ADC silicon area, and $P_{SUP}$ the power dissipation.

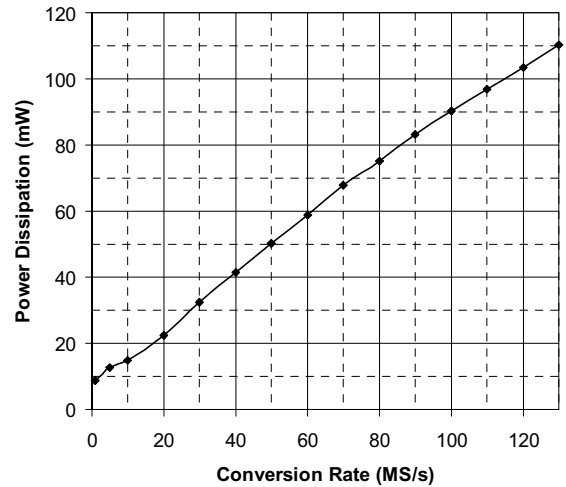

Fig. 4. Power dissipation versus conversion rate. The input frequency and signal swing is 10MHz and $2V_{P-P}$, respectively.

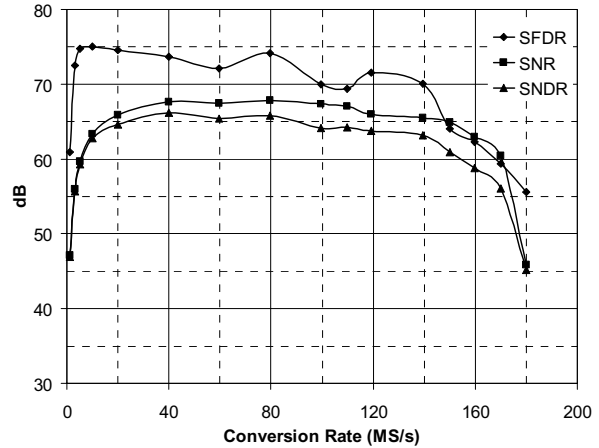

Fig. 5. SFDR, SNR, and SNDR versus conversion rate. The input frequency and signal swing is 10MHz and $2V_{P-P}$, respectively.

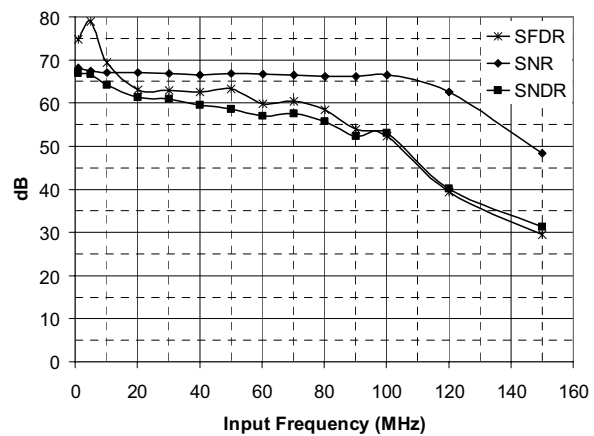

Fig. 6. SFDR, SNR and SNDR versus input frequency. The conversion rate and signal swing are 110MS/s and $2V_{P-P}$, respectively.

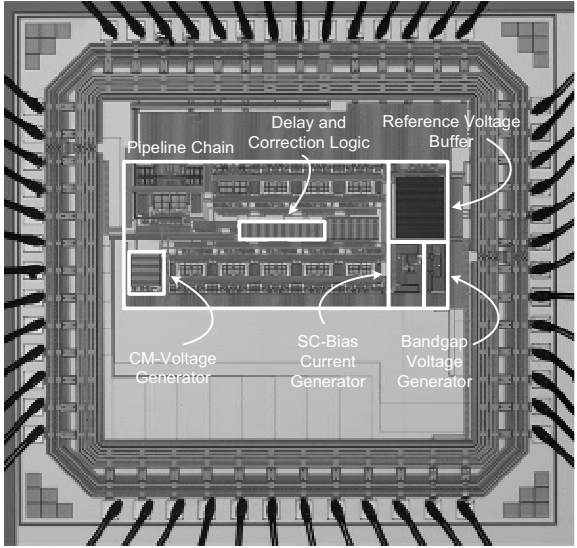

Fig. 7. Die photograph.

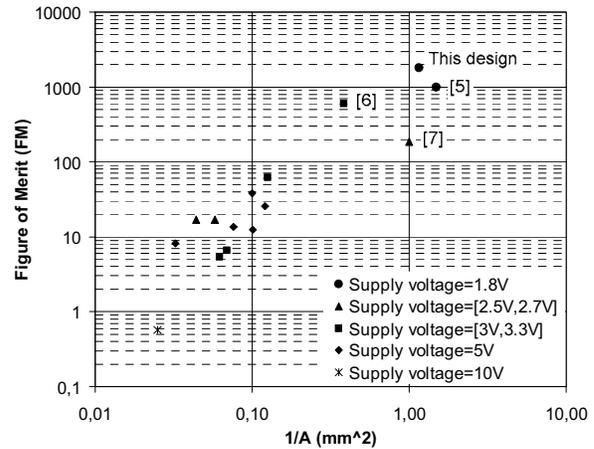

Fig. 8. Figure of Merit (*FM*) versus 1/*A* for 12b ADCs. $f_{CR}$ is given in MS/s, *A* is given in mm$^2$ and $P_{SUP}$ is given in mW.

|  | 110MS/s |
|---|---|
| Technology | 0.18μm digital CMOS |
| Nominal supply voltage | 1.8V |
| Resolution | 12 bit |
| Full Scale analog input | 2 $V_{P-P}$ |
| Area | 0.86 mm$^2$ |
| Analog Power Consumption | 97 mW |
| DNL | ±1.2 LSB |
| INL | -1.5/+1 LSB |
| SNR ($f_{in}$=10MHZ) | 67.1 dB |
| SNDR ($f_{in}$=10MHZ) | 64.2 dB |
| SFDR ($f_{in}$=10MHZ) | 69.4 dB |
| ENOB ($f_{in}$=10MHZ) | 10.4 bit |

Table I. Key data for the proposed 12b pipeline ADC.

$$FM = \frac{f_{CR} \cdot 2^{ENOB}}{A \cdot P_{SUP}} \quad (2)$$

Fig. 8 shows *FM* (2) versus the inverse of the silicon area of 15 12b ADCs including the ADC presented in this paper (named "This design"). The plot shows that this design has the highest *FM* and the 2$^{nd}$ lowest area consumption. Further, this converter is the 2$^{nd}$ published 12b ADC with 1.8V supply voltage. The ADCs [5]-[7] are closest in *FM* and also area consumption. The other ADCs in Fig. 8 are taken from *IEEE Proc. of ISSCC* and *IEEE Symposium on VLSI Circuits Digest of Technical Papers* over the last 9 years.

## 5.  Conclusions

This paper presents a pipeline ADC with 12 bits of resolution and 110MS/s nominal conversion rate. The effective number of bits is 10.4, the power consumption is 97mW and the silicon area is 0.86mm$^2$. The ADC utilizes a SC bias current generator that scale the bias current, and subsequently the power consumption, automatically as a function of the conversion rate. Low area and low and scalable power dissipation results in an ADC IP-block that is well suited for a wide range of application in SoC systems.

### References

[1] B. Hernes, A. Briskemyr, T. N. Andersen, F. Telstø, T. E. Bonnerud, Ø. Moldsvor, "A 1.2V 220MS/s 10b Pipeline ADC Implemented in 0.13μm Digital CMOS," *Proc. of ISSCC2004*, pp. 256-257, 2004.

[2] T. Byunghak Cho, P. R. Gray, "A 10 b, 20 Msample/s, 35 mW Pipeline A/D Converter," IEEE *Journal of Solid-State Circuits*, Vol. 30, pp. 166-172, March 1995.

[3] D. Kelly, W. Yang, I. Mehr, M. Sayuk, L. Singer, "A 3V 340mW 14b 75MSPS CMOS ADC with 85dB SFDR at Nyquist," *Proc. of ISSCC2001*, pp. 134, 2001.

[4] R. H. Walden, "Analog-to-Digital Converter Survey and Analysis," *IEEE Journal on Selected Areas in Communications*, vol. 17, no. 4, pp. 539-550, April 1999.

[5] A. Zjajo, H. Ploeg, M. Vertregt, "A 1.8V 12bits 80MSample/s Two-step ADC in 0.18-μm CMOS," *Proc. of ESSCIRC2003*, 2003.

[6] S. Kulhalli, V. Penkota, R. Asv, "A 30mW 12b 21MSample/s Pipelined CMOS ADC," *Proc. of ISSCC2002*, 2002.

[7] H. Ploeg, G. Hoogzaad, H. A. H. Termer, M. Vertregt, R. L. J. Roovers, "A 2.5V 12b 54MSampel/s 0.25um CMOS ADC in 1mm$^2$," *Proc. of ISSCC2001*, 2001.